\documentclass[apjl]{emulateapj}
\usepackage{psfig,epsf}
\shorttitle{The dwarf galaxy DDO 47: testing triaxial halos}
\shortauthors{Gentile et al.}

\begin{document}


\title{The dwarf galaxy DDO 47 as a dark matter laboratory: \\
testing cusps hiding in triaxial halos}


\author{G. Gentile\altaffilmark{1},
A. Burkert\altaffilmark{2},
P. Salucci\altaffilmark{1},
U. Klein\altaffilmark{3},
F. Walter\altaffilmark{4}
\email{gentile@sissa.it}
}

\altaffiltext{1}{SISSA-ISAS, Via Beirut 2, 34135 Trieste, Italy}
\altaffiltext{2}{Universit\"at M\"unchen, Scheinerstr. 1, 81679 M\"unchen, Germany}
\altaffiltext{3}{Radioastronomisches Institut der Universit\"at Bonn, Auf dem H\"ugel 71, 53121 Bonn, Germany}
\altaffiltext{4}{Max Planck Institut f\"ur Astronomie, K\"onigstuhl 17, 69117 Heidelberg, Germany}



\begin{abstract}
We present an analysis of high resolution HI data of the dwarf galaxy DDO 47, aimed at testing
the hypothesis that dark halo triaxiality might induce non-circular motions resulting
in rotation curves best fitted by cored halos, even if the dark matter halo is intrinsically
cuspy. This hypothesis could be invoked in order to reconcile the predictions of the standard
$\Lambda$CDM theory with the rotation curves of disk galaxies.
DDO 47 is an ideal case to test this hypothesis because it has a very regular velocity 
field, its rotation curve is best fitted by a cored halo and a NFW halo is inconsistent
with the data. 
We analysed the velocity field through the higher-order 
harmonic terms in order to search for kinematical signatures
of alleged non-circular motions needed to ``hide'' a cusp: the result is that
globally non-circular motions are at a level of 2-3 km s$^{-1}$, and 
they are more likely to be associated
to the presence of some spiral structure than to a global elongated potential 
(e.g. a triaxial halo).
These non-circular motions are far from being sufficient to 
account for the discrepancy with the $\Lambda$CDM predictions.
We therefore conclude that the dark matter halo around 
the dwarf galaxy DDO 47 is truly cored and that a cusp cannot be hidden by non-circular motions.
\end{abstract}





\section{Introduction}

A fundamental prediction of the cosmological ($\Lambda$) Cold Dark Matter ($\Lambda$CDM) theory is
that virialized dark matter halos should have universal density profiles.
An empirical fit formula for the spherically averaged
density distribution $\rho (r)$ had been proposed by Navarro, Frenk \& White
(1996, 1997, NFW)

\begin{equation}
\rho(r) = \frac{\rho_s}{(r/r_s)(1+r/r_s)^2}
\end{equation}

\noindent where $\rho_s$ and $r_s$ are two in principle free but for
a given cosmological model strongly correlated parameters. This density
distribution converges to power-law functions at small radii
($\rho \sim r^{-1}$) and at large radii ($\rho \sim r^{-3}$).
Moore et al. (1999) later on argued that
the inner cusp may even be steeper with $\rho \sim r^{-1.5}$.

The power-law behaviour has been a matter of strong debate.
Recent simulations, e.g. by Power et al. (2003) and Reed et al. (2005)
demonstrate that the CDM halo density profiles are better represented by a function
with a continuously varying slope and no convergence to
a power-law slope at small radii. A revised fitting
formula was derived by Navarro et al. (2004), where the slope at the 
innermost radius is about -1.2 for spiral galaxies and -1.35 for dwarfs. 
Merritt et al. (2005)
showed that the S\'ersic (1968) law fits high-resolution $\Lambda$CDM halos with a wide range
of masses with a S\'ersic index of $n \approx 2.4 - 3$ which is usually used
as a good fit of the surface brightness profiles of
the centres of early-type galaxies.

However, as we can typically trace rotation curves 
observationally in the regime of $0.1 r_s \leq r \leq 2 r_s$, the NFW profiles still 
provide a good description to the dark matter halos in galaxies.

The CDM predictions have been confronted with observations,
investigating the rotation curves of disk galaxies.
Numerous mass models of spiral, dwarf and LSB galaxies have shown that
in a large number of cases the inferred cores of dark matter halos
are much shallower than expected from the NFW-fit
(Flores \& Primack 1994, Moore 1994, Burkert 1995,
Salucci \& Burkert 2000, \citealt{dB:01}, \citealt{Sa:01}, \citealt{dBB:02}, 
\citealt{W:03}, Gentile et al. 2004). The cored density distribution
(Burkert 1995)

\begin{equation}
\rho(r) = \frac{\rho_0}{(1+r/r_c)(1+r/r_c)^2}
\end{equation}

\noindent fits the data often very well, indicating that, in contrast
to cosmological predictions, the maximum density of dark halos is finite.
Donato et al. (2004) even found a strong correlation between the core radius
$r_c$ and the dynamical mass indicating again that the observed cored
dark matter halos represent a 1-parameter family.

This contradiction between theoretical  predictions and
observations has stimulated a lot of discussions
as it has the potential to provide interesting new insights into
the nature of dark matter and its possible interaction with visible
matter (for reviews see Primack 2004 or Ostriker \& Steinhardt 2003).
In this letter we will explore the possibility 
that solid-body rotation curves which are usually interpreted as a signature 
of constant density cores would arise naturally also in NFW halos as a 
result of a triaxial dark matter
mass distribution that hides the steep central density cusp
inducing deviations from axial symmetry in the disk (e.g. Hayashi et al. 2004);
see however Gnedin et al. 2004 and Kazantzidis et al. 2004 for the effect of
baryons reducing halo triaxiality.
The deviations from axial symmetry in the gas 
kinematics can be tested observationally with galaxies showing 
two-dimensional velocity fields that are clearly inconsistent with
a NFW profile.

\section{The target: DDO 47}

An ideal case is the nearby dwarf 
galaxy DDO~47 for which HI observations (Walter \& Brinks 2001) 
with adequate resolution and sensitivity exist; these data show that
the kinematics of the gas is very regular, with a well-behaved
velocity field (Fig. 1). 

DDO 47 is one of the most clear cases with kinematical properties that
are inconsistent with NFW.
A first rotation curve of DDO 47 was 
presented in Walter \& Brinks (2001) and a more detailed decomposition,
performed by Salucci, Walter \& Borriello (2003), shows that 
a NFW halo is inconsistent with the data. They also show that the halo of DDO~47
has a central constant-density core: the rotation
curve was successfully fitted
with a Burkert halo with $r_c$=7 kpc and $\rho_0 = 1.4 \times 10^{-24}$
g cm$^{-3}$ (assuming a distance of 4 Mpc, and 1 arcmin corresponds
to 1.16 kpc);
any other cored halo (e.g. pseudo-isothermal) gives 
equally good fits to the rotation curve. We show the Burkert fit in Fig. 2,
where the rotation curve was derived by leaving the position angle as a free
parameter, in order to better trace the warp. The error bars are 
derived from the difference between the two sides, also considering a minimum
error of half the velocity resolution, corrected by the inclination.     
Also shown in Fig. 2 is the
NFW fit; differently from Salucci, Walter \& Borriello (2003) we do not fix
the virial mass $M_{vir}$ by matching the last velocity point, but we
leave it instead as a free parameter, the concentration parameter $c$ scaling
as $c=21(\frac{M_{vir}}{10^{11}\rm{M_{\odot}}})^{-0.13}$ (Wechsler et al. 2002).

Here we subject the 
same data to a more thorough analysis, carried out in the light of 
the above discussion. The spatial resolution of the datacube
($\sim$ 300~pc) is similar to that of the dwarf galaxy 
simulations performed by Navarro et al. (2004).

\begin{figure}
   \centerline{\psfig{figure=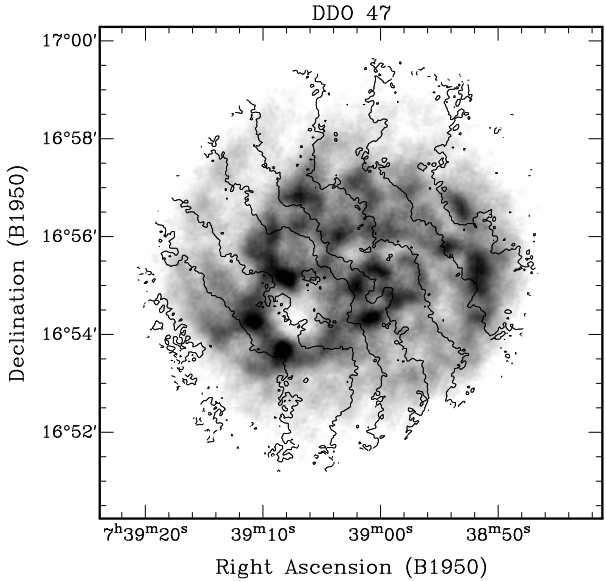,angle=-90,width=8.2cm}}
   \caption{
Velocity field of DDO 47 (contours) from the 15 $\arcsec$ resolution cube
and its total HI map (grey scale).
Contours are spaced by 10 km s$^{-1}$.
}
    \end{figure}

\begin{figure}
   \centerline{\psfig{figure=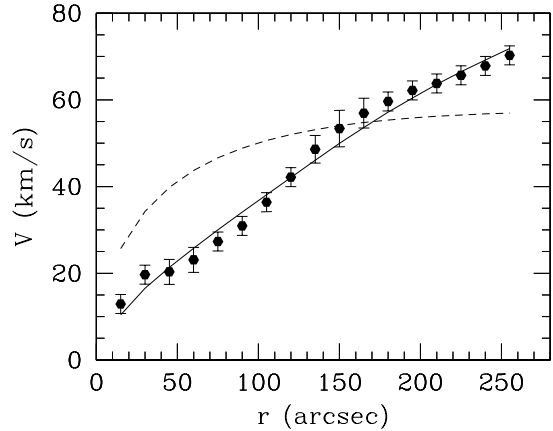,width=7.4cm}}
   \caption{
Mass models for DDO 47: the solid line is the best-fitting model with the Burkert halo, and 
the dotted line represents the best-fitting NFW model, with a virial mass $M_{vir}=2.4 \times
10^{10} M_{\odot}$ (see text). 
}
    \end{figure}

\begin{figure}
   \centerline{\psfig{figure=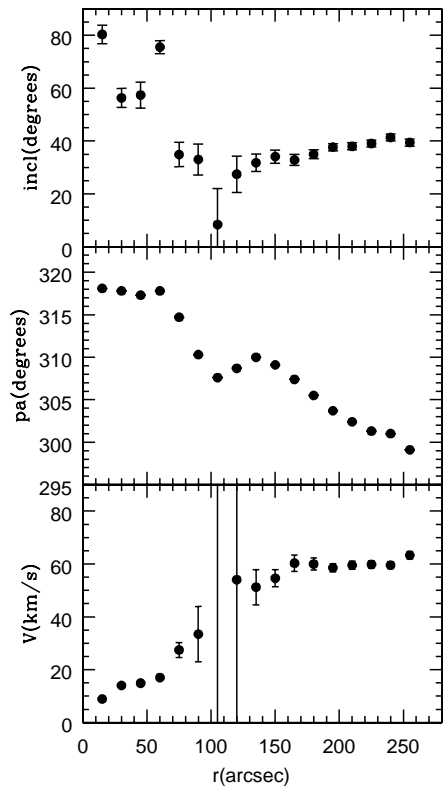,width=6.7cm}}
   \caption{
Tilted-ring fit of the velocity field with the inclination as a free parameter. From top
to bottom: inclination, position angle (fixed in this case, see text) and velocity 
as a function of radius. The uncertainties shown are formal errors.
}
    \end{figure}

\begin{figure}
   \centerline{\psfig{figure=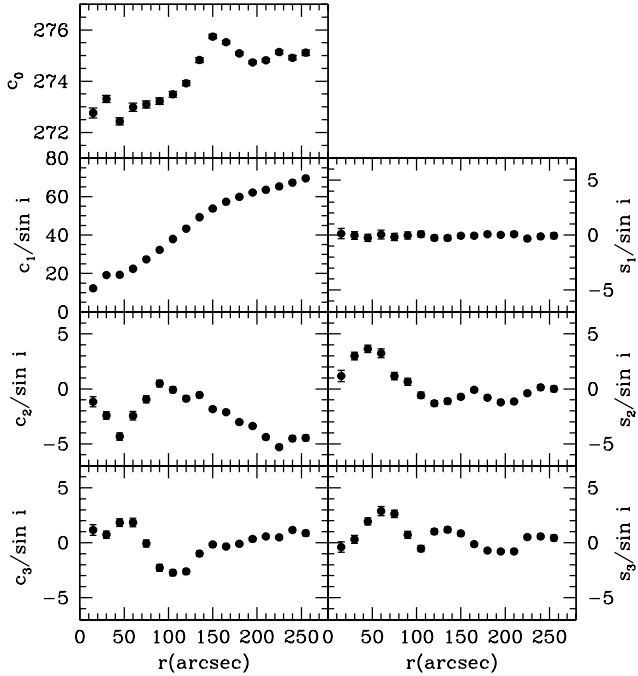,width=8.6cm}}
   \caption{
Harmonic decomposition of the velocity field, with terms up to the order 3. Notice
that here the coefficients of eq. 3 with $j \ge 1$ have been divided by ${\rm sin}~i$ in order to have
physical velocities instead of line of sight velocities. The uncertainties shown are formal errors.
}
    \end{figure}

We have made a tilted-ring analysis of the velocity field with the 
inclination and the centre fixed at the same values as used by Walter 
\& Brinks (2001). As noticed by these authors, leaving the inclination $i$
as a free parameter does not give a very stable result. In
order for a possible radial change of $i$ to account for the
discrepancy with the $\Lambda$CDM predictions, $i$ should increase as a function of
radius. Instead, as is shown in Fig. 3, 
the tilted-ring fit with the inclination free (with 35$^\circ$ as
initial guess) suggests that the radial
change of the inclination, if any, goes in the opposite direction 
(i.e. $i$ seems to get larger and not smaller for small radii).
Note that in this figure we have fixed the position angle at an average value
(separately for each radius) 
between fits made with different values of the inclination, between 5$^\circ$
and 65$^\circ$. Therefore, considering that there is no evidence for an
increase of the inclination, we conclude that fixing the inclination at 
35$^\circ$ is a reasonable assumption for the aim of this paper. We have also 
found that a slight change in the systemic velocity (273~km~s$^{-1}$ instead of 
272~km~s$^{-1}$) decreases the differences between the two sides.

\section{Harmonic decomposition, non-circular motions and halo triaxiality}

In order to better investigate the kinematic properties of DDO 47, we have fitted 
the velocity field in terms of harmonic coefficients, where the observed velocity
along the line of sight $V_{\rm los}$ is given by:

\begin{equation}
V_{\rm los} = c_0 + \sum_{j=1}^n [c_j {\rm cos}(j \psi) + s_j {\rm sin}(j \psi)]
\end{equation}

where $\psi$ is the azimuthal angle defined as the angle from the major axis of 
the receding side in a counterclockwise direction. In this analysis we only considered 
terms up to $j=3$. In the case of a rotating disk plus axisymmetric radial
motion (i.e. the traditional tilted-ring fit of the velocity field), only terms up to $j=1$ 
are present and $c_0=V_{\rm sys}$ (the systemic velocity), 
$c_1=V_{\rm rot}~ {\rm sin}~ i$ (the rotation velocity, and $i$ is the inclination angle), 
and $s_1=V_{\rm rad}~ {\rm sin}~ i$ (the radial velocity). 
More detailed descriptions of this
method and its applications can be found in Schoenmakers, Franx and de Zeeuw (1997),
Schoenmakers (1999) and Wong, Blitz \& Bosma (2004).

\begin{figure}
   \centerline{\psfig{figure=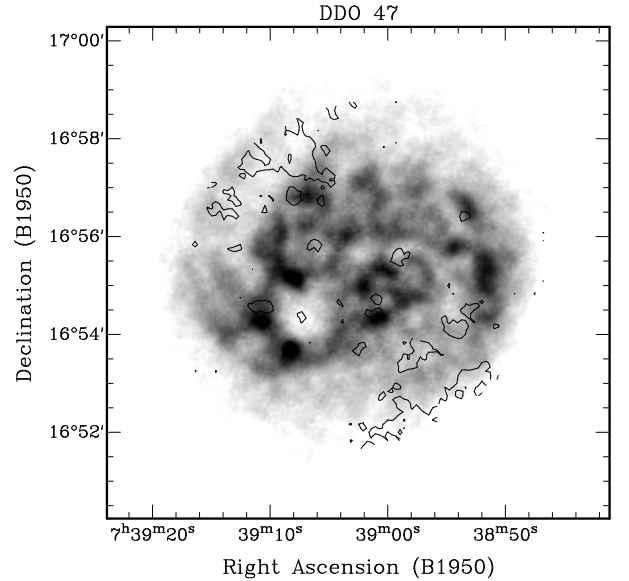,angle=-90,width=8.2cm}}
   \caption{
Residuals of the tilted-ring fit. Contours are -15, -10, -5, 5, 10, and 15
km s$^{-1}$.
}
    \end{figure}

\begin{figure}
   \centerline{\psfig{figure=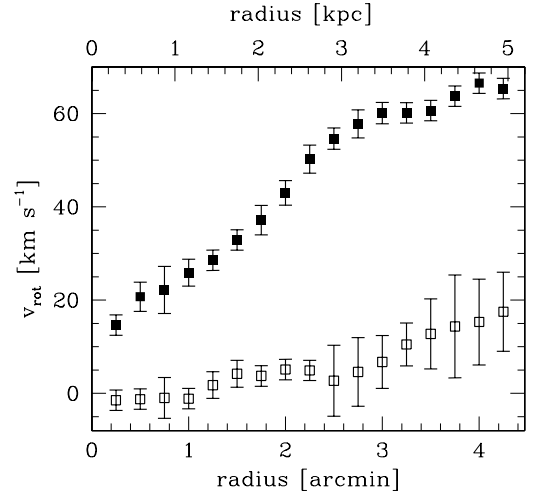,width=7.4cm}}
   \caption{
Observed rotation curves of the dwarf galaxy DDO~47 with the position angle fixed, along the major axis 
(filled squares) and along the minor axis (open squares). The stellar 
exponential scale length is 0.5 kpc.
}
    \end{figure}

The results of the harmonic decomposition of the velocity field are shown 
in Fig. 4, where the parameters were fitted with a fixed inclination (35$^{\circ}$) and  
the position angle left as a free parameter. First of all it can be noticed that the
$s_1$ term is negligible and that the $s_3$ term is larger but still has a small
amplitude, at most 3 km s$^{-1}$. The fact that $s_3/s_1 >> 1$ is to be expected,
as DDO 47 has a low inclination and this fraction is a strong function of inclination
(Schoenmakers 1999). As shown by Schoenmakers et al. (1997), a global elongation of the
potential would cause the $s_3$ term to be approximately constant and non-zero, while 
wiggles could be explained by the presence of spiral structure. In DDO 47 the latter
case applies, we find no evidence for a global elongation of the potential and 
the amplitude of the oscillations is much smaller (a factor 5-10) than the discrepancy
with the $\Lambda$CDM predictions. Fig. 4 also shows that some lopsidedness is present,
since the $j=2$ terms are non-zero. In the residual velocity field (Fig. 5) we find that 
the regions with residuals larger than 5 km $^{-1}$ (9 km $^{-1}$ deprojected)
are not a common feature.
The innermost ones are likely to be associated to either star-forming regions or to the possible bar,
while the outermost ones are probably due to asymmetry in the outer warp. 

As can be seen in Walter \& Brinks (2001), the major axis of the stellar body seems to be a bit 
misaligned with the HI disk, and low-level non-circular motions associated to the optical part of
the galaxy can be seen in the velocity field. Unfortunately due to the large number of stars in 
the field a more thorough analysis of the possible bar structure in the optical
image is not possible. If the $s_3 \sim 3$ km s$^{-1}$ region around 
60-70$\arcsec$ were associated to a barred potential, its amplitude would anyway be
much lower than needed to reconcile the rotation curve with the $\Lambda$CDM predictions.
We have discussed so far two of the three effects that according to Rhee et al. (2004) could
bias the rotation curve (the inclination and the bar). Concerning the third effect, i.e. the
influence of a possible bulge, these authors claim that the random motions of the bulge material
could bias the observed rotation velocities towards the systemic velocity. However in DDO 47
visual inspection of the velocity profiles shows that they are symmetric. This was further assessed
by building a velocity field where the velocity at any point is given by the peak of the profile:
if the profiles were strongly asymmetric, the difference between this peak velocity field and a
first-moment velocity field would be large and the separation between approaching and receding
side should be evident. Instead, we see small deviations and no obvious separation between the
two sides. 

We can therefore conclude that the non-circular motions in DDO 47 are at a level
of at most 3 km s$^{-1}$ and that they are not associated with a global elongation
of the potential. As a particular case in Fig. 6 we show the minor axis kinematics of DDO 47:
the velocity shows very small deviations from zero; the analysis performed earlier
gives a more general result, making use of the whole velocity field to derive the
amount of non-circular motions. We conclude that the global discrepancy between observations
and $\Lambda$CDM predictions shown in Fig. 2 cannot be explained by non-circular motions
induced by halo triaxiality, regardless of the viewing angle.

\section{Conclusions}

We have presented a dynamical analysis of HI data of the dwarf galaxy DDO 47 
to test the hypothesis that the possible triaxial shape of the dark matter
halo might induce deviations from circular motions resulting
in a rotation curve best fitted by a cored halo, with the dark matter halo being still
CDM-like (i.e., cuspy). 
The high-quality HI observations of
DDO 47 make it an ideal case to test this hypothesis because it has a very regular velocity 
field, its rotation curve is best fitted by a cored halo and a NFW halo is inconsistent
with the data (Salucci et al. 2003). 
Even though our analysis indicates a cored dark matter halo profile for DDO 47, 
work to test the effect of different halo shapes and disk/halo alignments
on the shapes of rotation curves is clearly very valuable.

We have performed a harmonic decomposition of the velocity field in order to 
estimate the amount of non-circular motions. We find that they are at a level
of at most 3 km s$^{-1}$ and that they are likely to be associated to spiral
structure rather than to a global elongation of the potential. In any case
the observed deviations from circular motion are much smaller than the discrepancy 
between $\Lambda$CDM predictions and the observed rotation curve.
We conclude therefore that the dark matter halo around 
the dwarf galaxy DDO 47 is truly cored and that a cusp cannot be hidden by non-circular motions.

\begin{acknowledgements}
GG wants to thank E. Hayashi and O. Valenzuela for useful conversations.
\end{acknowledgements}

\label{lastpage}

\begin{thebibliography}{99}


\bibitem[\protect\citeauthoryear{Burkert}{1995}]{B:95} Burkert, A., 1995, ApJ, 447, L25
\bibitem[\protect\citeauthoryear{de Blok, McGaugh \& Rubin}{2001}]{dB:01} de Blok, W. J. G., McGaugh, S. S., Rubin, V. C., 2001, AJ, 122, 2396
\bibitem[\protect\citeauthoryear{de Blok \& Bosma}{2002}]{dBB:02}
 de Blok, W. J. G., Bosma, A., 2002, A\&A, 385, 816
\bibitem[\protect\citeauthoryear{Donato et al.}{2004}]{Do:95} Donato, F., Gentile, G., \& Salucci, P.\ 2004, MNRAS, 353, L17 
\bibitem[\protect\citeauthoryear{Flores \& Primack}{1994}]{Fl:94} Flores, R., Primack, J.R., 1994, ApJ, 427, L1
\bibitem[\protect\citeauthoryear{Gentile et al.}{2004}]{Ge:04} Gentile, G., Salucci, P., Klein, U., Vergani, D., Kalberla, P., 2004, MNRAS, 351, 903
\bibitem[\protect\citeauthoryear{Gnedin et al.}{2004}]{Gn:04} Gnedin, O.~Y., Kravtsov,  A.~V., Klypin, A.~A
., \& Nagai, D.\ 2004, ApJ, 616, 16
\bibitem[\protect\citeauthoryear{Hayashi et al.}{2004}]{Ha:04} Hayashi, E., Navarro, J. F., Jenkins, A., Frenk, C. S., Power, C., White, S. D. M., Springel, V., Stadel, J., Quinn, T., Wadsley, J., 2004, astro-ph/0408132
\bibitem[\protect\citeauthoryear{Kazantzidis et al.}{2004}]{Ka:04} Kazantzidis, S., Kravtsov, A. V., Zentner, A. R., Allgood, B., Nagai, D., Moore, B., 2004, ApJ, 611, L73
\bibitem[\protect\citeauthoryear{Merritt et al.}{2005}]{Me:05} Merritt, D., Navarro, J.~F., Ludlow, A., \& Jenkins, A.\ 2005, ApJL, 624, L85 
\bibitem[\protect\citeauthoryear{Moore}{1994}]{Mo:94} Moore, B., 1994, Nat., 370, 629
\bibitem[\protect\citeauthoryear{Moore et al.}{1999}]{Mo:99} Moore, B., Quinn, T., Governato, F., Stadel, J., Lake, G., 1999, MNRAS, 310, 1147
\bibitem[\protect\citeauthoryear{Navarro et al.}{1996}]{NFW:96} Navarro, J. F., Frenk, C. S., White, S. D. M., 1996, ApJ, 462, 563
\bibitem[\protect\citeauthoryear{Navarro et al.}{1997}]{NFW:97}Navarro, J.~F., Frenk, C.~S., \& White, S.~D.~M.\ 1997, ApJ, 490, 493 
\bibitem[\protect\citeauthoryear{Navarro et al.}{2004}]{Na:04}Navarro, J. F., Hayashi, E., Power, C., Jenkins, A. R., Frenk, C. S., White, S. D. M., Springel, V., Stadel, J., Quinn, T. R., 2004, MNRAS, 349, 1039
\bibitem[\protect\citeauthoryear{Ostriker \& Steinhardt}{2003}]{OS:03}Ostriker, J.~P., \& Steinhardt, P.\ 2003, Science, 300, 1909 
\bibitem[\protect\citeauthoryear{Power et al.}{2003}]{Po:03} Power, C., Navarro, J.~F., Jenkins, A., Frenk, C.~S., White, S.~D.~M., Springel, V., Stadel, J., \& Quinn, T.\ 2003, MNRAS, 338, 14 
\bibitem[\protect\citeauthoryear{Primack}{2004}]{Pr:04} Primack, J.~R.\ 2004, IAU Symposium, 220, 53 
\bibitem[\protect\citeauthoryear{Reed}{2005}]{Re:05} Reed, D., Governato, F., Verde, L., Gardner, J., Quinn, T., Stadel, J., Merritt, D., \& Lake, G., 2005, MNRAS, 357, 82 
\bibitem[\protect\citeauthoryear{Rhee et al.}{2004}]{2004} Rhee, G., Valenzuela, O.,  Klypin, A., Holtzman
, J., \& Moorthy, B.\ 2004, ApJ, 617, 1059
\bibitem[\protect\citeauthoryear{Salucci \& Burkert}{2000}]{SB:00} Salucci, P., Burkert, A., 2000, ApJ, 537, L9
\bibitem[\protect\citeauthoryear{Salucci}{2001}]{Sa:01} Salucci, P., 2001, MNRAS 320, L1
\bibitem[\protect\citeauthoryear{Salucci et al.}{2003}]{Sa:03} Salucci, P., Walter, F., \& Borriello, A.\ 2003, A\&A, 409, 53 
\bibitem[\protect\citeauthoryear{Schoenmakers, Franx \& de Zeeuw}{Schoenmakers et al.}{1997}]{Sch:97} Schoenmakers, R. H. M., Franx, M., de Zeeuw, P. T., 1997, MNRAS, 292, 349
\bibitem[\protect\citeauthoryear{Schoenmakers}{1999}]{Sch:99} Schoenmakers, R. H. M., Ph.D. thesis, Univ. Groningen 
\bibitem[\protect\citeauthoryear{S\'ersic et al.}{1968}]{Se:68} S\'ersic, J. L. 1968, Atlas de Galaxies Australes (C\'ordoba: Obs. Astron., Univ. Nac. C\'ordoba)
\bibitem[\protect\citeauthoryear{Walter \& Brinks}{2001}]{Wa:01} Walter, F., \& Brinks, E.\ 2001, AJ, 121, 3026 
\bibitem[\protect\citeauthoryear{Wechsler et al.}{2002}]{W:02} Wechsler, R. H., Bullock, J. S., Primack, J. L., Kravtsov, A. V., Dekel, A., 2002, ApJ, 568, 52
\bibitem[\protect\citeauthoryear{Weldrake, de Blok \& Walter}{Weldrake et al.}{2003}]{W:03} Weldrake, D.T.F., de Blok, W. J. G., Walter, F., 2003, MNRAS, 340, 12
\bibitem[\protect\citeauthoryear{Wong, Blitz \& Bosma}{Wong et al.}{2004}]{Wo:04} Wong, T., Blitz, L., Bosma, A., 2004, ApJ, 605, 183
\end{thebibliography}
\end{document}